\documentclass{JHEP3}
\usepackage{amsmath,amssymb}
\usepackage{units}
\usepackage{slashed}
\usepackage[koi8-r]{inputenc}
\usepackage[russian,english]{babel}
\usepackage{graphicx}

\newcommand{\lsim}{\lesssim}
\newcommand{\gsim}{\gtrsim}

\newcommand{\Tr}{\mathop\mathrm{Tr}}

\newcommand{\bk}{\mathbf{k}}
\newcommand{\be}{\begin{equation}}
\newcommand{\ee}{\end{equation}}
\newcommand{\ba}{\begin{eqnarray}}
\newcommand{\ea}{\end{eqnarray}}
\renewcommand{\l}{\left(}
\renewcommand{\r}{\right)}

\newcommand{\half}{\frac{1}{2}}

\newcommand{\cL}{\mathcal{L}}

\newcommand{\e}{\mathrm{e}}
\newcommand{\const}{\mathop\mathrm{const}}
\newcommand{\nuMSM}{\ifmmode\nu\mathrm{MSM}\else$\nu$MSM\fi}

\newcommand{\Xcr}{X_{\mathrm{cr}}}

\preprint{LMU-ASC 26/11}

\title{Late and early time phenomenology of Higgs-dependent cutoff}
\author{F. Bezrukov\\
  Arnold Sommerfeld Center for Theoretical Physics,\\
  Department f\"ur Physik, Ludwig-Maximilians-Universit\"{a}t
  M\"{u}nchen,\\
  Institute for Nuclear Research of the Russian Academy of Sciences,\\
  60th October Anniversary prospect 7a, Moscow 117312, Russia\\
  E-mail: \email{Fedor.Bezrukov@physik.uni-muenchen.de}}
\author{D. Gorbunov\\
  Institute for Nuclear Research of the Russian Academy of Sciences,\\
  60th October Anniversary prospect 7a, Moscow 117312, Russia\\
  E-mail: \email{gorby@ms2.inr.ac.ru}
}
\author{M. Shaposhnikov\\
  Institut de Th\'eorie des Ph\'enom\`enes Physiques,\\
  \'Ecole Polytechnique F\'ed\'erale de Lausanne,\\
  CH-1015 Lausanne, Switzerland    \\
  E-mail: \email{Mikhail.Shaposhnikov@epfl.ch}
}

\abstract{
  The analysis of theories with non-minimal coupling of Higgs field to
  gravity revealed that they enter into strong coupling regime above
  certain Higgs-dependent cutoff, which may be considerably below the
  Planck scale.  Assuming that the effective theory, complementing the
  Standard Model or its minimal extension---the $\nu$MSM---contains a
  set of higher dimensional operators suppressed by the
  Higgs-dependent cutoff, we analyse the reheating of the Universe
  after the Higgs inflation.  We show that extra terms do not spoil
  the Higgs inflation, but can lead to baryogenesis and to warm
  sterile neutrino dark matter production at the reheating stage of
  the Universe expansion.  They can also result in neutrino mass
  generation and proton decay.}


\makeatletter
\def\bbl@cite@choice{%
  \global\let\bibcite\org@bibcite%
  \global\let\bbl@cite@choice\relax}
\makeatother

\begin{document}

\section{Introduction}
\label{intro}

The Universe is flat, homogeneous and isotropic, and has the
primordial density perturbations with almost scale-invariant
spectrum.\footnote{It is well known for a long time, e.g.:
  \foreignlanguage{russian}{<<\emph{Чудище обло, озорно, огромно с
      тризевной и Лаей,>>} В.~Тредиаковский, \emph{Телемахида} (1766).}}
All these above may be well explained by a hypothetical inflationary
stage of the Early Universe before it became hot.  This stage can be
realized within the Standard Model (SM) of particle physics extended
by non-minimal coupling of the SM Higgs field $H$ to gravity
\cite{Bezrukov:2007ep}
\begin{equation}
  \label{HI}
  \cL_{\mathrm{gravity}} = \frac{M_P^2}{2}R
  + \xi H^\dagger{}H R
  \;, 
\end{equation}
where $M_P\equiv 1/\sqrt{8\pi G_N}=\unit[2.44\times10^{18}]{GeV}$ is
the reduced Planck mass.   In \cite{Bezrukov:2008ut} (see also
\cite{GarciaBellido:2008ab}) we analyzed the history of the Universe
within the Higgs inflation scenario in two theories, assuming their
validity up to the Planck scale.  The first one was the Standard Model
and the second one was the $\nu$MSM (an extension of the SM by 3
singlet fermions with masses below the electroweak scale, for a review
see ref.~\cite{Boyarsky:2009ix}).  One of our main goals in
\cite{Bezrukov:2008ut} was to determine the initial conditions for the
hot Big Bang.  In other words, we addressed the question whether the
quantum numbers, effectively conserved in the SM (such as $B-L$, where
$B$ and $L$ are baryon and lepton numbers respectively) and in the
$\nu$MSM (primordial abundances of the singlet leptons) can be
generated during inflation or reheating of the Universe.  For this end
we added to the SM and to the $\nu$MSM higher dimensional operators,
breaking these conservation laws, and suppressed by the scale $\zeta
M_P$. The bounds on parameter $\zeta$ were derived from requirement
that the Higgs inflation is not spoiled.  The constraints on $\zeta$
allowed to estimate the $B-L$ excess and the abundances of singlet
fermions after reheating.  We found that the contribution of higher
dimensional operators, bounded in this way, to the production of Warm
Dark Matter (WDM) and to baryogenesis is negligible.  We concluded,
therefore, that the production of sterile neutrino WDM\footnote{On the
contrary, a sterile neutrino with the mass exceeding 100 keV (a Cold
Dark Matter candidate) can be created during the reheating stage
of the Universe in the necessary amount.} and baryon asymmetry of the
Universe must be a low-temperature phenomenon, having nothing to do
with inflation and reheating.

The aim of the present paper is to revisit this analysis in connection
with new results obtained recently in the investigations of
consistency of Higgs inflation.  We present below a short account of
this progress, to motivate our present study.

For successful inflation yielding primordial density perturbations
with the amplitude matched to the cosmological observations
\cite{Komatsu:2008hk}, it is required at the tree level that the
non-minimal coupling $\xi$ must be rather large,
\begin{equation}
  \label{eq:xiCOBE}
  \xi\simeq47000\sqrt{\lambda}
  \;,
\end{equation}
where $\lambda$ is the Higgs field self-coupling constant.  This
relation is modified by loop corrections, see
\cite{Bezrukov:2008ej,Bezrukov:2009db,DeSimone:2008ei,Barvinsky:2009ii}
for numerical results and \cite{Bezrukov:2010jz} for further
discussion.\footnote{Everywhere in this paper, except for the Section
\ref{sec:LowEnergy}, we will need the values of the constants at the
inflationary scale.  Then, this relation is hardly modified compared
to the tree level result, and the only subtlety is the connection of
the Higgs self-coupling at high scale with the physical Higgs mass.
Under the assumptions, formulated in ref.~\cite{Bezrukov:2010jz},
this relation is quite precisely given by the SM renormalization group
running, see Fig.~5 of ref.~\cite{Bezrukov:2009db}.}

The requirement $\xi\gg 1$ raised doubts in the validity of the Higgs
inflation \cite{Burgess:2009ea,Barbon:2009ya}.  The estimate of the
high energy scattering amplitudes in the \emph{low energy electroweak
vacuum} showed that the theory enters into a strong-coupling regime (the
tree unitarity is broken) at energies $E$ well below the Planck scale,
$E \simeq M_P/\xi$.  Since this energy is much smaller than the values
of the Higgs field during inflation, $H \sim M_P/\sqrt{\xi}$, the
authors of the papers \cite{Burgess:2009ea,Barbon:2009ya}  concluded
that the Higgs inflation is an ``unnatural'' phenomenon, as it
intrinsically requires fine-tuning of the Higgs potential at large
values of the Higgs field.

However, the analysis \cite{Bezrukov:2010jz} (see also
\cite{Ferrara:2010yw}) revealed that the theory remains in a weak
coupling regime below certain cutoff scale $\Lambda(h)$, which
\emph{in general depends on the Higgs field} and is in fact
\emph{above} the inflationary scale.  This makes the Higgs inflation
self-consistent and ``natural'' \cite{Bezrukov:2010jz,Ferrara:2010yw}.
However, the phenomenology and predictivity of the model depend on the
properties of the UV-complete theory at high energies
\cite{Bezrukov:2010jz}.  Here we analyze the post-inflationary
phenomenology using the effective field theory parametrization of the
effects which might come from the high energy theory.

One can distinguish several cutoff scales corresponding to violation
of perturbative unitarity in the different sectors of the model.  In
the gravity-scalar sector the cutoff is\footnote{We present all
cutoffs in the Jordan frame, corresponding to eq.~(\ref{HI}) and use
the unitary gauge, $H^T=\l 0 \,, h/\sqrt{2}\r$.}
\begin{equation}
  \label{scalescalar}
  \Lambda_{g-s}\l h\r \simeq \left\{
    \begin{array}{l@{\;,\quad\text{for }}l}
      \frac{M_P}{\xi}    & h\lesssim \frac{M_P}{\xi} \;, \\
      \frac{\xi\,h^2}{M_P} & \frac{M_P}{\xi}\lesssim h
                           \lesssim \frac{M_P}{\sqrt{\xi}}
                           \;, \\
      \sqrt{\xi}{h}      & h\gtrsim \frac{M_P}{\sqrt\xi} \;.
    \end{array}
  \right.
\end{equation}
The cutoff for purely gravitational interactions (graviton scattering
off graviton) is just the effective Planck scale (see eq.~(\ref{HI}))
\begin{equation}
  \label{scalePlanck}
  \Lambda^2_{\mathrm{Planck}} \simeq M_P^2+\xi h^2
  \;.
\end{equation}
The cutoff associated with SM gauge interactions at 
$h>\frac{M_P}{\xi}$ is
\begin{equation}
  \label{scalegauge}
  \Lambda_{\mathrm{gauge}} \simeq  h
  \;.
\end{equation}
and coincides with $\frac{M_P}{\xi}$ at smaller fields $h$.
\FIGURE{
  \includegraphics[width=0.5\textwidth]{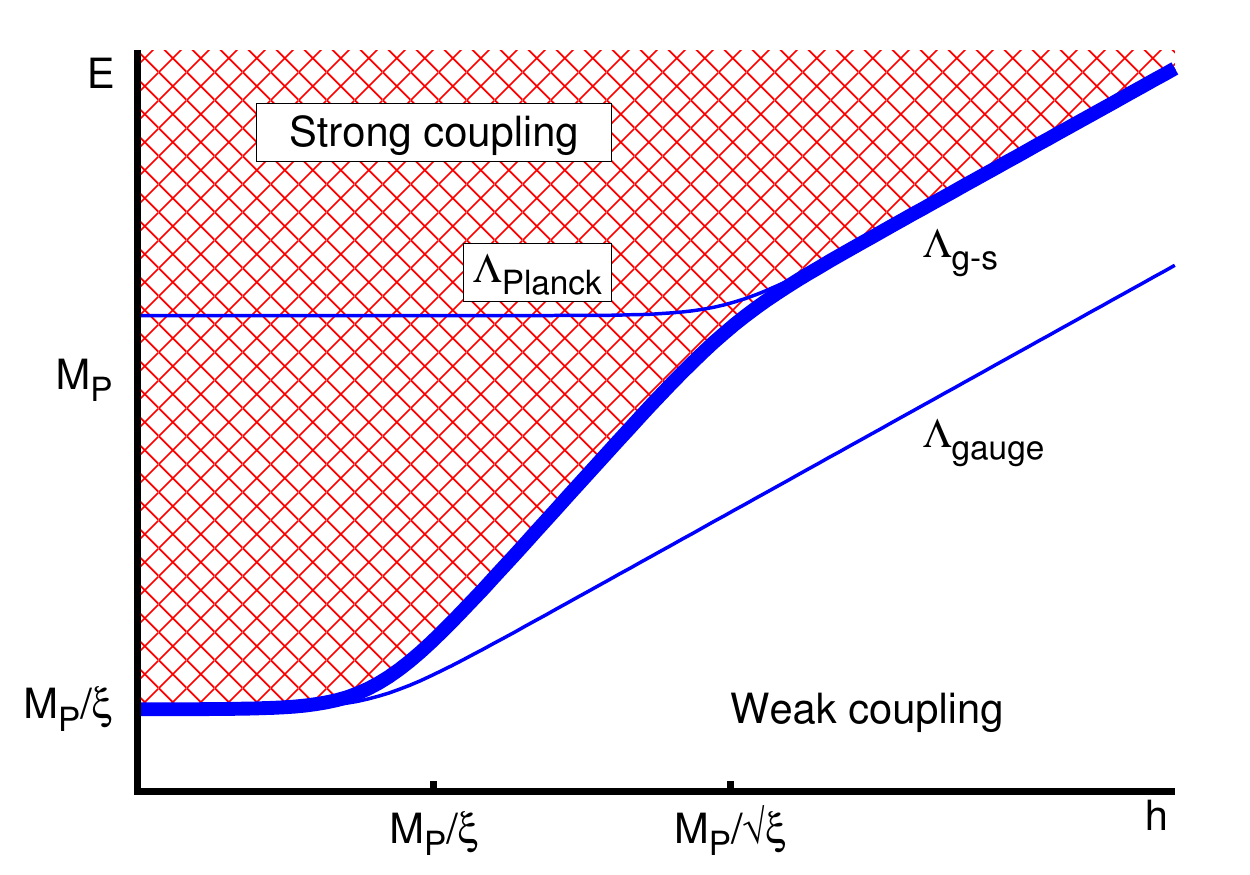}%
  \includegraphics[width=0.5\textwidth]{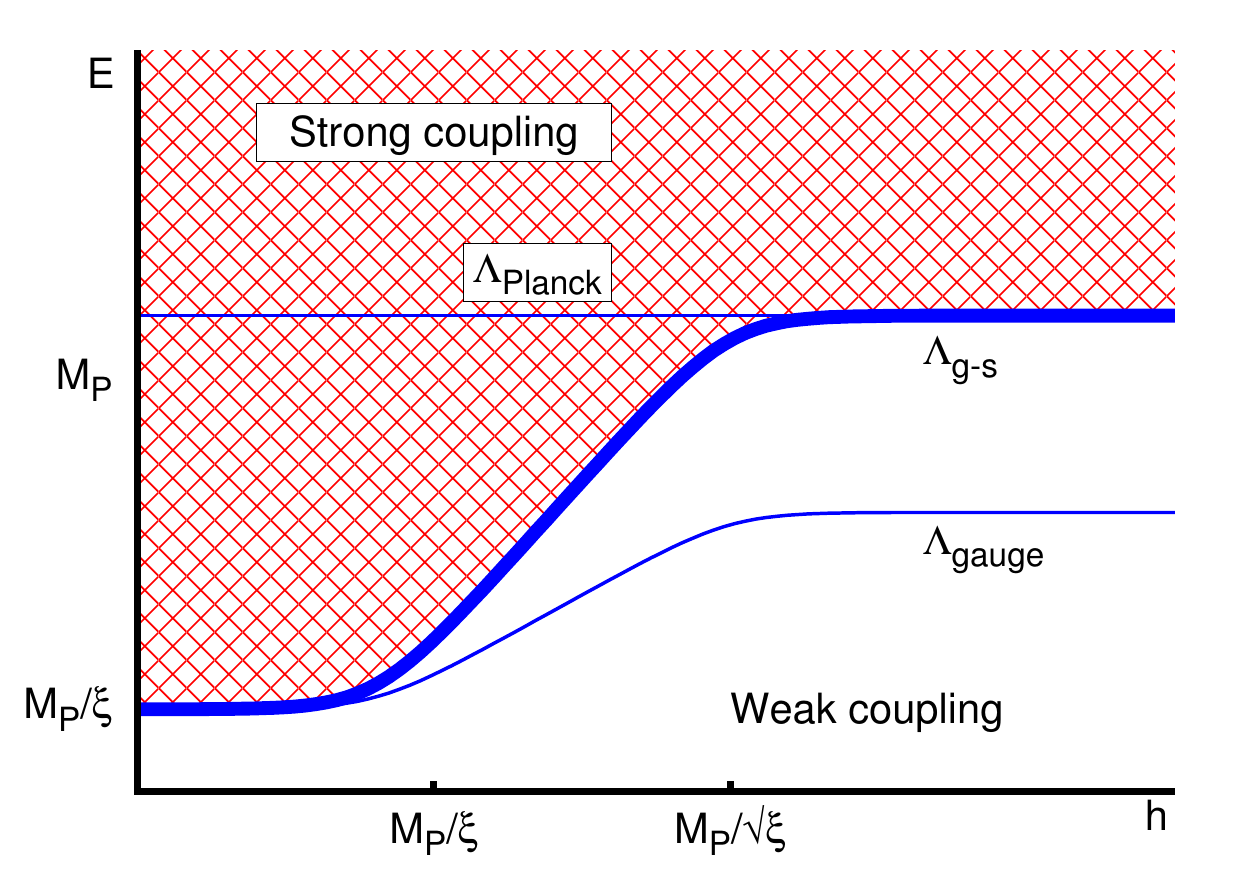}
  \caption{Schematic depiction of the cut-offs (\ref{scalescalar}),
    (\ref{scalePlanck}), and (\ref{scalegauge}) in the Jordan (left)
    and Einstein (right) frames.}
}

One of the reasons the inflation takes place while the theory remains 
in weakly coupling regime  is because it does not involve large
momentum transfers.  The validity of the theory above momenta
$\Lambda(h)$ is unclear. Thus, the question arises what happens when
momentum transfers $E$ are between $\Lambda_{\mathrm{gauge}}\l h\r$ and
$\Lambda_{\mathrm{Planck}}$.  Remaining within the minimal
setup,\footnote{Alternatively, it is possible to attempt perturbative
  UV completion by introducing new states at the cut-off
  scale \cite{Giudice:2010ka}.}
we can imagine two scenarios:
\begin{enumerate}
\item The Lagrangian of gravity + SM (actually, its viable and
  phenomenologically complete extension such as the $\nu$MSM) remains
  valid up to the Planck scale (without any higher order operators
  suppressed by lower scales), but the system at some $h$ and large
  momenta $\Lambda_{\mathrm{gauge}}\l h\r<E<\Lambda_{\mathrm{Planck}}$ is in the
  strong coupling regime and requires non-perturbative methods for its
  description. Essentially this point of view was taken in our
  previous studies of inflation and reheating of the Universe
  \cite{Bezrukov:2007ep,Bezrukov:2008ut,Bezrukov:2008ej,Bezrukov:2009db}.
\item The consistency of the theory in this range of energies requires
  introduction (below the cut-off scale) of higher order operators,
  suppressed by \emph{a Higgs-dependent cutoff $\Lambda\l h\r$}.
\end{enumerate}

The present paper is devoted to the study of phenomenology of the
second scenario in the SM and $\nu$MSM, as the first one was
considered in \cite{Bezrukov:2008ut}.  The fact that the cutoff is
considerably smaller than the Planck scale may induce a number of
interesting effects at low energies, such as proton decay or neutrino
mass generation.  In the early Universe, the baryon and lepton number
violating reactions are greatly enhanced in comparison with the first
scenario, which can potentially change baryogenesis or leptogenesis.
The sterile neutrino WDM primordial production at the reheating stage
after inflation may rise considerably in the second scenario.  At the
same time, even with the small, but \emph{Higgs-dependent} cutoff, the
inflation \emph{is not modified.}

The paper is organized as follows.  In Section
\ref{Sec:high-order-operators} we discuss higher-dimensional operators
which could emerge in the model with Higgs-dependent cutoff
\eqref{scalescalar} and explain why they do not spoil the Higgs
inflation.  The case of the SM is considered in Section \ref{Sec:SM}.
We study low-energy phenomenology and reconsider the processes in
post-inflationary Universe.  We show that higher-dimensional operators
may produce lepton asymmetry and hence be responsible for the
matter-over-antimatter domination in the present Universe.  We explore
the possible role of higher-dimensional operator in reheating. Section
\ref{Sec:nuMSM} is entirely devoted to the $\nu$MSM and we address
here all the questions we do in Section \ref{Sec:SM}.  We show, in
addition, that the model new fields---sterile neutrinos---might form
warm dark matter, and describe new options for reheating and
leptogenesis.  We summarize our results in Section \ref{summary},
concluding with discussion of UV-completion of the model with regard
to cosmology.

\section{Prerequisite discussion: 
Higgs-inflation framework and higher dimensional operators}
\label{Sec:high-order-operators}

We start with introduction of relevant framework and short review on
the relevant facts about the Universe evolution in the model with the
non-minimally coupled Higgs inflation following
refs.~\cite{Bezrukov:2007ep,Bezrukov:2008ut}.

From now on all the computations in the paper we are going to make in
the Einstein frame, which is connected to the Jordan frame (\ref{HI})
by the conformal transformation
\begin{equation}
  g_{\mu\nu}\to  \hat{g}_{\mu\nu} = \Omega^2 g_{\mu\nu}
  \;,\quad
  \Omega^2(h) = \frac{M^2+\xi h^2}{M_P^2}
  \;.
\end{equation}
This change induces modification of the
kinetic term of the scalar field, which can be returned to canonical
form by the field redefinition
\begin{equation}
  \label{eq:chi(h)}
  \chi \simeq \left\{
    \begin{array}{l@{\qquad\text{for}\quad}l}
      h
      & h<X_{\mathrm{cr}}\equiv\sqrt{\frac23} \frac{M_P}\xi
      \;,\\
      \sqrt{\frac{3}{2}}M_P\log \Omega^2(h)
      & h > X_{\mathrm{cr}}\equiv\sqrt{\frac23} \frac{M_P}\xi
      \;.
    \end{array}
  \right.
\end{equation}
The action (in the Einstein frame) then becomes an action of the 
usual minimally coupled to gravity scalar field $\chi$ with the
potential
\begin{equation}
  \label{eq:5}
  U(\chi) =
  \frac{1}{\Omega^4\left[ h \l \chi\r \right]}
  \frac{\lambda}{4}\left[h^2\l \chi\r -v^2\right]^2
  \;.
\end{equation}
The rest of the action (after proper rescaling of the fermion fields)
looks like the usual SM action in the unitary gauge with proper
rescaling of the mass terms by $\Omega\left[h\l\chi\r\right]$.  I.e.\
the usual SM mass terms for fermion $\psi$ and vector boson $A_\mu$ 
get the general form of
\begin{equation}
\label{rescaled-masses}
  \cL_{\mathrm{Yukawa}} = \frac{Y
    h(\chi)}{\Omega\left[h\l\chi\r\right]}
  \bar\psi\psi
  \;,\quad
  \cL_{\mathrm{gauge}} = \frac{g^2h^2(\chi)}{2\Omega^2\left[h\l\chi\r\right]}
  A_\mu A^\mu
  \;,
\end{equation}
with $Y$ and $g$ being the corresponding Yukawa and gauge coupling
constants.

The potential (\ref{eq:5}) provides for slow roll inflation in the
region $h>M_P/\sqrt{\xi}$ ($\chi>M_P$), and gives rise to primordial
density perturbations of observed amplitude for model parameters
obeying \eqref{eq:xiCOBE}.  For small field values $h<M_P/\xi$   and
equally small energies the model behaves as the  usual SM:  this is
the usual thermal evolution of the Universe. 

In the intermediate region $M_P/\xi<h<M_P/\sqrt{\xi}$  the scalar
potential is approximately quadratic. Oscillations of the scalar field
with frequency 
\begin{equation}
\label{oscillation-frequency}
\omega\equiv \sqrt{\frac{\lambda}{3}}\,\frac{M_P}{\xi}
\end{equation}
and amplitude in the intermediate region above reheat the Universe. 
\emph{In the absence of the higher dimensional  operators suppressed
by Higgs-dependent cutoff}  the reheating happens via production of
gauge and Higgs bosons by the oscillating field background as
described in \cite{Bezrukov:2008ut,GarciaBellido:2008ab}, and the
reheating temperature $T_r$ is confined in the range
\begin{equation}
  \label{eq:55}
  1.4\times 10^{-5}M_P
  < T_{r} <
  4.5\times10^{-5}  \l \frac{\lambda}{0.25}\r^{1/4}M_P
  \;,
\end{equation}
where the relation \eqref{eq:xiCOBE} is put in. 

The overall conformal factor $\Omega\l h\r$ becomes relevant only for
$h>M_P/\sqrt{\xi}$, which corresponds to the inflationary regime. This
means, that after inflation (starting from the post-inflationary
reheating) the value of the cut-off in the Jordan and Einstein frames
are the same.  As far as the current work deals mostly with the post
inflationary processes, the only change we need to make to get to the
Einstein frame with canonically normalized fields is the variable
change (\ref{eq:chi(h)}).

Let us proceed with the discussion on higher dimensional operators.  As
far as the full UV complete theory for the non-minimally coupled Higgs
inflation is not yet developed, \emph{one can not make any definite
predictions.}  Meanwhile we can explore what happens if various higher
dimensional operators are added to the model.

We will analyze the effect of the following higher dimensional 
terms 
\begin{subequations}
  \label{general-operators}
\begin{align}
 \label{general-operators-higgs}
  \delta\cL_\mathrm{NR} =
  & - \frac{a_6}{\Lambda^2}(H^\dagger H)^3 + \dotsb \\
 \label{general-operators-violating}
  & + \frac{\beta_L}{4\Lambda}F_{\alpha\beta}
      \bar{L}_\alpha \tilde{H}H^\dagger L^c_\beta
    + \frac{\beta_B}{\Lambda^2} O_{\text{baryon violating}}
    + \dotsb + \mathrm{h.c.} \\
 \label{general-operators-sterile}
  & + \frac{\beta_N}{2\Lambda} H^\dagger H \bar{N}^cN +
    \frac{b_{L_\alpha}}{\Lambda}\bar{L}_\alpha(\slashed{D}N)^c\tilde{H}
    + \dotsb
  \;,
\end{align}
\end{subequations}
where $L_\alpha$ are SM leptonic doublets, $\alpha=1,2,3$, $N$ stands
for right handed sterile neutrinos potentially present in the model,
$\tilde H_a=\epsilon_{ab}H_b^*$, $a,b=1,2$; here $\Lambda$ depends on
the background (Higgs) field value, $\Lambda=\Lambda\l h\r$.  

In \cite{Bezrukov:2008ut} the limits on these operators were obtained
in the assumption that the cut-off scale is constant.  Then these
operators either directly change the inflationary potential, like the
Higgs potential (\ref{general-operators-higgs}), either generate the
contributions to the potential by loop corrections.  For example, from
the first two operators in (\ref{general-operators}) contributions in
the Jordan frame are $h^6/\Lambda^2$ and $h^8/\Lambda^4
\log(m^2(h)/\mu^2)$.  Both terms grow much faster than $h^4$ and can
not be compensated by the fourth power of the conformal factor, thus
leading to non-flat potential and bad inflationary properties.
Numerical estimate \cite{Bezrukov:2008ut} shows, that the minimal
value of constant cut-off, consistent with inflationary predictions,
is $\Lambda_{\mathrm{const}}\sim M_P$.  However, all the choices
(\ref{scalescalar}), (\ref{scalePlanck}), and (\ref{scalegauge}) grow
proportionally to $h$ in the inflationary regime, and the dangerous
terms in the potential behave just as $h^4$ for inflation.  Then after
conformal transformation the shape of the inflationary potential
(\ref{eq:5}) remains the same, with modified constants.  This means,
that the inflationary predictions for the initial density
perturbations remain unchanged, and the only modification is a
correction to the normalization (\ref{eq:xiCOBE}) of the non-minimal
coupling.

Perturbativity up to energy scale $\Lambda$ implies that the numerical
coefficients $a$, $\beta$ are of order one or less.  With the lack of
knowledge about underlying theory, this is the only requirement on
higher dimensional terms \eqref{general-operators}.  However, one can
distinguish three different situations.

First, it could be that still unknown dynamics at the scale
$\Lambda\ll M_P$ does not violate itself any global symmetries of the
SM.  Then, numerical coefficients in front of corresponding
symmetry-violating operators are suppressed, so that the effective
energy scale entering, say, baryon number violating operators in
\eqref{general-operators} is the Plank mass or even higher.  The
opposite case is when all gauge invariant operators enter
\eqref{general-operators} with numerical coefficients of order one. 
Finally, the intermediate situation corresponds to some hierarchy in
these coefficients.  As we observe below, this third situation is the
most interesting, as such models exhibit the richest phenomenology. In
particular, we will discuss in due course the case where at the scale
$\Lambda$ higher order operators could dominate over similar ones of
lower order.  In particular, for the operators of the form
\begin{equation}
  \label{yy}
  \delta\cL^\tau = y_\tau L_\tau H E_\tau + \beta_y L_\tau H E_\tau
  \frac{H^\dagger H}{\Lambda^2}
  +\dotsb
\end{equation}
one can fancy the hierarchy
\begin{equation}
  \label{exotic-hierarchy}
  1\sim \beta_y \gg y_\tau \sim 10^{-2} \;.
\end{equation}
For small $H$ this gives nothing new.  However, during reheating we
can get $H\sim\Lambda$, so the second term in \eqref{yy} can overcome
the first one at large values of the Higgs field, and make a
significant contribution compared to the lowest order estimate. Note,
that hierarchy \eqref{exotic-hierarchy} \emph{does not generally imply
strong coupling below the Plank scale.} To simplify further discussion
we suppose that all similar higher order operators are negligible.

Note in passing that in this work we treat a particular set of
couplings $\beta,a_6,\dotsc$ from pure phenomenological point of view
without any relation to what happens to the operators
\eqref{general-operators} above the scale $\Lambda$ and without any
discussion of matching conditions at critical values of the Higgs
field \eqref{scalescalar}.  Having in mind a possible UV-completion
like asymptotic safety \cite{Weinberg1979}  we also disregard all
dangerous contributions naively expected in the theory
\eqref{general-operators} with a formal high-energy cutoff at the
scale $\Lambda$, such as quadratic in  $\Lambda$ corrections to the
Higgs boson mass and others.

\section{Higher dimensional operators in the SM}
\label{Sec:SM}

\subsection{Role of the non-renormalizable  operators in low energy
  (present day) physics}
\label{sec:LowEnergy}

At present time the Higgs field dependence of the cutoff in the
(\ref{general-operators}) is not observable, since the Higgs
background has relaxed to its vacuum value $h=v=\unit[246]{GeV}$.  So
the effect of the higher dimensional operators
\eqref{general-operators} can be analyzed as usual, and is most
important for the lowest values of the cutoff, that is $M_P/\xi$, see
eqs.~\eqref{scalescalar}--\eqref{scalegauge} and the nearby 
discussion.  As far as the operators are suppressed by this scale, the
experimentally interesting effects can be expected from the operators
violating  symmetries of the renormalizable SM action.  These are the
operators in \eqref{general-operators-violating} responsible for  the
violation of lepton and baryon numbers.

The lowest order non-renormalizable lepton number violating operator
(first term in \eqref{general-operators-violating}) provides
the Majorana mass terms for active neutrinos
\begin{equation}
  \label{5dim-neutrino-masses}
  \cL_{\nu\nu}^{(5)} =
  \frac{\beta_Lv^2}{4\Lambda}\frac{F_{\alpha\beta}}{2}
  \bar\nu_\alpha\nu^c_\beta
  + \mathrm{h.c.}
  \;.
\end{equation}
To explain the observed pattern of the active neutrino masses the scale
of this lepton number violating operator should not be higher than
\begin{equation}
  \label{neutrino-mass-limit}
  \Lambda<2.8\times \unit[10^{14}]{GeV} \times \beta_L \times
  \l \frac{\unit[3\times10^{-3}]{eV^2}}{\Delta m^2_{\mathrm{atm}}}
  \r^{1/2}
  \;.
\end{equation}
For the allowed range of parameter $\xi$ at the electroweak scale
(see \cite{Bezrukov:2009db}, Fig.~4 for discussion and relation to the
Higgs boson mass), $\xi\simeq 10^3-10^5$, we
get for the scale $\Lambda$ 
\begin{equation}
  \label{scale-at-EW}
  \Lambda=\frac{M_P}{\xi}\sim \unit[\l 0.3-35\r\times 10^{14}]{GeV}
  \;.
\end{equation}
The neutrino masses can be explained without taking too large
parameter $\beta$, namely with 
\begin{equation}
\label{beta-L}
\beta_L\simeq 0.1-10\;.
\end{equation}

The next operator, important at low energies, is the baryon number
violating one, as it leads to the proton decay. We can use the
estimate of the proton decay in the standard $SU(5)$ model to place a
lower limit on the scale $\Lambda$.  Indeed, for the $SU(5)$ case the
proton life-time is~\cite{Hisano:2000dg} (the exact form of the
relevant dimension-6 operators contributing to the decay is given in
the reference)
\begin{equation}
  \label{SU5-proton}
  \tau_{p\to \pi^0 e^+} = 1.0\times \unit[10^{35}]{years}
  \times
  \l \frac{1/25}{g_5^2/4\pi} \r^2
  \l \frac{M_V}{\unit[10^{16}]{GeV}} \r^4
  \;,
\end{equation}
where $M_V=g_5 v_5/2$, and $v_5$ is the vev responsible for
spontaneous breaking of $SU(5)$ gauge symmetry. Thus, current limit on
the proton   decay, $\tau_{p\to \pi^0 e^+} >1.6\times 10^{33}$\,years 
\cite{PDG2010},   yields a bound on the scale $v_5$, which in our case
corresponds to the scale $\Lambda$ so that 
\begin{equation}
  \label{proton-decay-limit}
  \Lambda \gtrsim \sqrt{\beta_B}\times \unit[10^{16}]{GeV}\,
  \times
  \l
    \frac{\tau_{p\to \pi^0 e^+}}{1.6\times \unit[10^{33}]{years}}
  \r^{1/4}
  \;.
\end{equation}

The discrepancy between the limits \eqref{neutrino-mass-limit} and
\eqref{proton-decay-limit} is about 40 (for $\beta_B=\beta_L=1$) and
can be reconciled only with a particular hierarchy between the
dimensionless coefficients: 
\begin{equation}
\label{B-L-hierarchy}
  \beta_L\gtrsim 36\times\sqrt{\beta_B}
  \;.
\end{equation}
For example, if $\beta_L\sim 1$ within the interval \eqref{beta-L}, 
one has to chose $\beta_B\lesssim 10^{-3}$.  Thus, baryon number
violating scale cannot be $\Lambda=M_P/\xi$, it is certainly higher.

This implies, that while the neutrino masses can be explained for a
wide range of the Higgs masses, if the lepton number is violated at
the scale (\ref{scalegauge}) or (\ref{scalescalar}), the proton decay
should be prevented by some means. If allowed, its rate must be 
suppressed by a higher scale, like gravity scale  \eqref{scalePlanck}.
The required hierarchy between $\beta_L$ and $\sqrt{\beta_B}$ though
large, still may be accidental; then it suggests that proton decay has
to be observed in the nearest future experiments. Otherwise, the
hierarchy may be a natural outcome of some unknown mechanism operating
at the scale of UV-completion. Indeed, for instance, the two operator
under discussion  \eqref{general-operators-violating} differ in a
particular respect: one involves the Higgs fields, while another does
not. A specific renormalization flow evolution of the Higgs-containing
operators might be responsible for establishing  the required
hierarchy \eqref{B-L-hierarchy} at the scale $M_P/\xi$.

\subsection{Neutrino reheating via dimension-5 operators}
\label{subsec:neutr-rehe-via}

It was shown in \cite{Bezrukov:2008ut}, that if the suppression scale
equals the Planck mass, then the particle production  by the higher
dimensional operators is negligible, and hence their impact on the
reheating.  However, this is  not immediately obvious with the
Higgs-field dependent cut-off. Let us analyze the contribution to
reheating of the lepton number violating operator in
(\ref{general-operators-higgs}).

\paragraph{Scalar gravity sector cutoff.} We start from the operator
\eqref{5dim-neutrino-masses}, and omit the
numerical coefficients $F_{\alpha\beta}\sim 1$ in further estimates. During the preheating,
while the background field evolves according to eq.~(3.4) of
\cite{Bezrukov:2008ut}, we have the following effective time-dependent
Majorana type mass source for the active neutrinos
\begin{equation}
  \label{eq:3}
  m(t) \simeq \left\{
   \begin{array}{l@{\;,\quad\text{for }}l}
    \frac{\beta_L}{4}\sqrt{\frac{3}{\lambda}}\omega
    & \chi(t)\gtrsim \sqrt{\frac{3}{\lambda}}\omega \;,\\
    \frac{\beta_L}{4}\sqrt{\frac{\lambda}{3}}\frac{1}{\omega}\chi^2
    & \chi(t)\lesssim \sqrt{\frac{3}{\lambda}}\omega \;.\\
   \end{array}
  \right.
\end{equation}
While the number of generated fermions of given 3-momentum ${\bf k}$ 
is small, $n_\bk \ll 1$, the effect of this can be treated by first
order perturbation theory.  The calculation described in Appendix
\ref{app:nugen_scalargravity} yields for the fermion density 
saturated by the moment of reheating 
\begin{equation}
  \label{eq:11-bis}
  n = {\beta_L^2} \frac{\sqrt{\lambda}}{16\sqrt{3}\pi^2}
  \xi\Xcr^3
  \;.
\end{equation}

Without a hierarchy in $F_{\alpha\beta}$  the total number density of
all three species of active neutrinos  is roughly trice the amount of 
(\ref{eq:11-bis}).  To check, whether this leads to preheating, we
should compare the energy density in neutrinos with that in the
inflaton oscillating with amplitude $X$, see  Appendix
\ref{app:nugen_scalargravity} for details. The equality between these
two culminates the reheating, and the corresponding reheating
temperature would be 
\begin{equation}
  \label{eq:18-bis}
  T_r \sim \left(\frac{10\,\lambda}{27\,g_*}\right)^{1/4}
  \frac{\beta_L}{2\pi\sqrt{2\pi}}\frac{M_P}{\sqrt{\xi}}
  \;.
\end{equation}
Though it is parametrically larger than the temperature of reheating 
due to renormalizable operators \cite{Bezrukov:2008ut},
$T_r^{\mathrm{renorm}}\propto \frac{M_P}{\xi}$, but numerically, for
$\beta_L\sim 1$, it is of the same order, cf. \eqref{eq:55}
\[
  T_r \simeq 1.7\times \beta_L \times
  \unit[10^{14}]{GeV} \sim T_r^{\mathrm{renorm}}\simeq
  \unit[10^{14}]{GeV}
  \;.
\]
Hence, higher order operators do not significantly enhance reheating.

We check in Appendix \ref{app:nugen_scalargravity} that the Pauli
blocking is relevant for $\beta_L\sim 1$ if the reheating temperature
is not at the lower end of the allowed range \eqref{eq:55}. This
diminishes the possible contribution of higher dimensional operators
to reheating: they are never really relevant to this process.  

\paragraph{Gauge sector cutoff.} One could think, that with the lower
cut-off (\ref{scalegauge}) the contribution to reheating becomes
significant.  However, with more active generation Pauli blocking
effects would suppress the process much earlier, so the contribution
of the generated neutrinos to the energy balance during reheating can
not really exceed calculations in the previous section.  Of course, as
before, one would expect that still existence of additional reheating
channel somewhat favors  the higher end of the reheating temperature
interval \eqref{eq:55}. 

\bigskip
Finally, we conclude, that the estimate~(\ref{eq:55})
is still a valid one, with additional
preference for the higher temperature part of the allowed region.

\subsection{Lepton asymmetry in the Early Universe from 5-dimensional
  operator}
\label{subsec:lepton-asymmetry}

The next interesting effect which can appear from the lepton violating
operators in (\ref{general-operators}) is the generation of the baryon
asymmetry via leptogenesis.

One can ask whether it is possible to generate a reasonably large
lepton asymmetry $\Delta n_L\equiv n_L-n_{\bar L}$ with the
nonrenormalizable operator  \eqref{general-operators-violating} at
the  post-inflationary preheating stage.  At a given time the amount of
asymmetry is determined by averaging of lepton number operator $\hat Q
_L$, which evolves in accordance with the Heisenberg equation,
\[
i\,\frac{d}{dt}\, \hat Q _L = \left[\hat H_{\mathrm{int}} ,\hat Q _L\right]\;.
\]

Here the interaction Hamiltonian comprises both SM interactions and
all nonrenormalizable terms \eqref{general-operators}. From these
equations one finds in a usual way (see, e.g.\
\cite{Shaposhnikov:1987tw}),
that at a given time the production rate of
lepton asymmetry, $d\Delta n_L/dt$ is proportional to the imaginary
part of the minimal nontrivial trace of the SM Yukawa couplings
$Y_{\alpha\beta}=Y_\alpha\cdot \delta_{\alpha\beta}$ entering
\begin{equation}
  \label{lepton-yukawas}
  \cL_Y = -Y_\alpha\bar L_\alpha H E_\alpha + \mathrm{h.c.}
  \;,
\end{equation}
and to the couplings $F_{\alpha\beta}$ entering 
\eqref{general-operators-violating}.
This trace reads
\begin{equation}
  \label{CP-violating-trace}
  \beta_L^4\Tr \l FF^\dagger F Y Y F^\dagger Y Y \r
  \;,
\end{equation}
and, since the Yukawa constants $Y_{\alpha\beta}$ can be chosen to be 
diagonal, the dominant term is
\begin{equation}
  \label{Yukawas-dominant}
  \beta_L^4 Y_3^4 F_{3\beta}F_{\alpha\beta}^* F_{\alpha3} F_{33}^*
  \;,
\end{equation}
where $Y_3=y_\tau$ obviously provides with the largest contribution.
For simplicity we adopt the unitary gauge while making an estimate of
the generated asymmetry.   The Majorana terms
\eqref{general-operators-violating} connect only neutrinos, while the
Dirac terms \eqref{lepton-yukawas}, instead, connect electrically
charged leptons.  Thus, to get the term \eqref{CP-violating-trace}
from a corresponding fermionic loop one has to insert the electroweak
vertices with $W$-bosons everywhere between  the vertices proportional
to $Y$ and $F$ (overall four electroweak vertices).  These $W$-bosons
can be either virtual or external.  The former case implies 3-loop
contributions, while the latter case actually describes scattering of
$W$-bosons produced in the early Universe via Higgs field oscillations
as explained in ref.~\cite{Bezrukov:2008ut}.

Let us consider the 3-loop contribution, which is numerically
suppressed by the loop factor
\begin{equation}
  \label{loop-factors}
  \l \frac{1}{4\pi^2}\r^3\sim 10^{-5}
  \;.
\end{equation}
During the reheating the relevant energies for the generation
asymmetry are given by the Higgs field oscillation frequency
$\omega$, which is much smaller than the gauge boson mass at this
epoch, see \eqref{rescaled-masses}, \eqref{oscillation-frequency}
\[
  \omega\equiv \sqrt{\frac{\lambda}{3}}\,\frac{M_P}{\xi} 
  \ll M_W \equiv g\sqrt{\frac{M_P\,\left|\chi\right|}{2\sqrt{6}\,\xi}}
  \;.
\]
Thus, $W$-bosons can be actually integrated out which gives the effective
Fermi lagrangian with the following combination playing the role of
the squared Fermi constant (see eq.~(3.8) in \cite{Bezrukov:2008ut})
\begin{equation}
  \label{W-boson-factors}
  \frac{g^4}{M_W^4}\sim \frac{\xi^2}{M_P^2\chi^2} \sim \frac{1}{h^4}
  \;.
\end{equation}
The couplings \eqref{general-operators-violating} and
\eqref{lepton-yukawas} yield the Higgs-dependent factor (see eq.~(3.9)
in \cite{Bezrukov:2008ut})
\begin{equation}
  \label{loop-factors-3++}
  \frac{\chi^2 M_P^2}{\xi^2} \cdot \l \frac{\chi M_P}{\xi\Lambda}\r^4
  \sim h^4\left(\frac{h^2}{\Lambda}\right)^4
  \;.
\end{equation}
Collecting the factors (\ref{loop-factors}), (\ref{Yukawas-dominant}),
(\ref{W-boson-factors}), and (\ref{loop-factors-3++}) we get for the
asymmetry generated during one oscillation of the Higgs field
\[
  \Delta n_L^{(1)} \sim \omega^{-1}\cdot \l \frac{1}{4\pi^2}\r^3
  \cdot \l \frac{m_\tau}{v} \r^4 \cdot \frac{\omega^4}{\lambda^2}\cdot
  \l \frac{\chi}{\Lambda}\r^4 \beta_L^4
  \sim
  \omega^{-1}\l \frac{1}{4\pi^2}\r^3
  y_\tau^4\beta_L^4 h^4\left(\frac{h}{\Lambda}\right)^4
  \;.
\]
Let us now estimate the result for the possible choices of the
high-energy cutoff scale.

\paragraph{Gauge cutoff.} The maximal effect is expected from the
lowest available cutoff, the gauge one (\ref{scalegauge}).  In this
case the ratio $h/\Lambda$ goes to one, and asymmetry generation rate
scales as $\Delta n_L^{(1)}\propto h^4\propto \chi^2$.  The amplitude
$X$ of the oscillations of the field $\chi$ is inversely proportional
to the cosmological time.  This means, that integration of the
asymmetry over a period of matter dominated expansion corresponds to
multiplication by the time of expansion (or number of oscillations).  
Recall, we found in Section \ref{subsec:neutr-rehe-via}  that
reheating develops here almost as it does without any higher
dimensional operators.   For reheating culminating at the amplitude of
the Higgs field oscillations $X_r$ (see \cite{Bezrukov:2008ut}) the
number of field oscillations during reheating is $N\sim \xi\Xcr/X_r$,
and the entropy after reheating is $s\sim
g_*^{1/4}\lambda^{3/4}(X_r\Xcr)^{3/2}$.  Collecting this we get for
the lepton asymmetry
\begin{equation}
  \label{L-asymmetry-naive}
  \Delta_L \simeq \frac{\Delta n_L^{(1)} N}{s}
  \sim \beta_L^4 y_\tau^4 \l \frac{1}{4\pi^2}\r^3
  \frac{1}{g_*^{1/4}\lambda^{5/4}}\xi
  \left(\frac{\Xcr}{X_r}\right)^{1/2}
  \;.
\end{equation}
For the estimated in \cite{Bezrukov:2008ut} range of reheating, caused
by lower order operators, eq.~\eqref{L-asymmetry-naive} 
gives the asymmetry to be in the range
\[
  \beta_L^4 \left(\frac{y_\tau}{0.01}\right)^4
  \left(\frac{0.25}{\lambda}\right)^{7/4} \times 10^{-15} \times\xi
  < \Delta_L <
  \beta_L^4 \left(\frac{y_\tau}{0.01}\right)^4
  \left(\frac{0.25}{\lambda}\right)^{3/2} \times 5\times10^{-15} \times\xi
  \;,
\]
or, inserting relation \eqref{eq:xiCOBE}, 
\begin{equation} 
\label{LNV}
  \beta_L^4 \left(\frac{y_\tau}{0.01}\right)^4
  \left(\frac{0.25}{\lambda}\right)^{5/4} \times 10^{-10}
  < \Delta_L <
  \beta_L^4 \left(\frac{y_\tau}{0.01}\right)^4
  \left(\frac{0.25}{\lambda}\right) \times 10^{-9}
  \;,
\end{equation}
where the smaller value corresponds to the larger reheating
temperature.  One can see, that if $\beta_L\sim1$ this may be just on
the very edge of reasonable value. In particular, for the ranges
\eqref{LNV}, \eqref{beta-L} one can both explain the active neutrino
masses and generate a sufficient amount of  lepton asymmetry at
preheating stage.  However, the result is rather sensitive to the
higher dimensional operators in the SM lepton sector (\ref{yy}).  As
far as during reheating $h\sim\Lambda$, all the higher order operators
in (\ref{yy}) are not automatically suppressed compared to the leading
one.  This means, that just the largest one contributes (assuming that
the whole sum still behaves nicely).  Thus, for the example of
hierarchy \eqref{exotic-hierarchy} the result
(\ref{L-asymmetry-naive}) is enhanced by a factor $(\beta_y/y_\tau)^4$
leading to increase in the net lepton asymmetry up to 8 orders of
magnitude. In this case quite large amount of lepton asymmetry can be
produced.  Note, that the higher order operators enhancing the lepton
asymmetry generation are obviously irrelevant for lepton number
violating processes at low energies discussed in Section
\ref{sec:LowEnergy} and hence phenomenologically viable. 

\paragraph{Gravity-scalar sector cutoff.} Otherwise, if we adopt the
gravity-scalar sector cutoff  \eqref{scalescalar}, which is  the
smallest one available, then the ratio $h^2/\Lambda$ goes to
$M_P/\xi$. We are left with $\Delta n_L^{(1)}\propto\const$, so in the
expanding Universe the asymmetry is saturated at lower times, that is
at the reheating.   Integration of the asymmetry over a period of
matter dominated expansion corresponds\footnote{With accuracy of a
factor of order one, which depends also on the treatment of
field-dependent cutoff in the calculation.} to multiplication of
$\Delta n_L^{(1)}$ by the total number of oscillations. Then, for
reheating which happened at the amplitude of the Higgs field
oscillations $X_r$ (see \cite{Bezrukov:2008ut}), it took approximately
$N\sim \xi\Xcr/X_r$ oscillations and lead to the entropy after
reheating $s\sim g_*^{1/4}\lambda^{3/4}(X_r\Xcr)^{3/2}$.  Collecting 
all the relevant factors we get for the lepton asymmetry
\begin{equation}
  \label{L-asymmetry-naive-2}
  \Delta_L \simeq \frac{\Delta n_L^{(1)} N}{s}
  \sim \frac{\Delta n_L^{(1)}}{g_*^{1/4}\lambda^{3/4}}\frac{\xi}{\Xcr^3}
  \left(\frac{\Xcr}{X_r}\right)^{5/2}\sim 
\beta_L^4 Y_\tau^4 \l \frac{1}{4\pi^2}\r^3
  \frac{2\sqrt{2}}{3\,g_*^{1/4}\lambda^{5/4}}\xi
  \left(\frac{\Xcr}{X_r}\right)^{5/2}
  \;.
\end{equation}
For the estimated in \cite{Bezrukov:2008ut} range of reheating, caused
by lower order operators, this means asymmetry in the range
\[
  \beta_L^4 \left(\frac{0.25}{\lambda}\right)^{15/4} \times 2.7\times 10^{-17} \xi
  < \Delta_L <
  \beta_L^4 \left(\frac{0.25}{\lambda}\right)^{5/2} \times10^{-14} \xi
  \;,
\]
or adopting \eqref{eq:xiCOBE} 
\begin{equation}
\label{asymmetry-range}
  \beta_L^4 \left(\frac{0.25}{\lambda}\right)^{13/4} \times 6.3\times 10^{-13}
  < \Delta_L <
  \beta_L^4 \left(\frac{0.25}{\lambda}\right)^2 \times 2.4\times 10^{-10}
\end{equation}
where smaller value corresponds to larger reheating temperature.  One
can see, that if $\beta_L\sim1$, this is too small asymmetry.
However, the result is rather sensitive to the higher dimensional
operators in the SM lepton sector (\ref{yy}).  By the end of reheating
we have $H^\dagger H /\Lambda^2\simeq 3 X_{\mathrm{cr}}/4X_r$ and all the
operators in (\ref{yy}) contribute proportional to their coupling
constants.  Thus, dimension-6 operator \eqref{yy} with hierarchy
\eqref{exotic-hierarchy} still leads to enhancement, but not that
large as with the gauge cutoff, and decreasing with higher reheating
temperatures.  The result (\ref{L-asymmetry-naive}) is enhanced by a
factor $(3\beta_y X_{\mathrm{cr}}/4y_\tau X_r)^4$ leading to increase in
the net lepton asymmetry: lower and upper limits of
\eqref{asymmetry-range} grow by factors of $15$ and $2\times10^5$,
respectively.  We should keep in mind, that the higher value of the
reheating temperature is more realistic (especially with account of
additional contributions from neutrino reheating described in Section
\ref{subsec:neutr-rehe-via}), and at high reheating temperatures even
this enhancement does not lead to viable leptogenesis.  So, generation
of sufficient baryon asymmetry with scalar-gravity cutoff is possible,
but may be problematic, unless $\beta_L$ takes highest values in the
interesting interval \eqref{beta-L}. 

\vskip 0.7cm 
Finally, at later stages of the Universe evolution the first term in
\eqref{general-operators-violating} gives rise to lepton violating
scattering in plasma \cite{Fukugita:1990gb}, which might wash out the
lepton asymmetry generated at preheating.  To clear up the situation
we adopt formulas from ref.~\cite{Nelson:1990ir} for our case of
$\nu_\tau$ playing the major role in lepton asymmetry generation. 
With parameters in front of the first term in
\eqref{general-operators-violating} yielding the correct values of
masses and mixing angles in neutrino sector we found that the lepton
violating scatterings are slower than the Universe expansion rate at
$T\gtrsim \unit[10^{14}]{GeV}$ and hence are out of equilibrium.  This
temperature is about the reheating temperature of the model
\eqref{eq:55}, and the washing out processes (if any) are not
efficient.  At later times the sphaleron processes come into
equilibrium and transfer the lepton asymmetry generated at preheating
to the net baryon asymmetry \cite{Khlebnikov:1996vj} $\Delta_B\approx
\Delta_L/3$.  From the estimates above, \eqref{LNV},
\eqref{asymmetry-range} we conclude that the baryon asymmetry of the
Universe can be explained with higher dimensional operators, if
suppressed by the gauge cut-off (\ref{scalegauge}), and with some
hierarchy like \eqref{exotic-hierarchy} this conclusion becomes 
certain. In the case of gravity-scalar cutoff \eqref{scalescalar} the
hierarchy like \eqref{exotic-hierarchy}  is definitely required in
order to produce the sufficiently large amount of the baryon
asymmetry. Moreover, the reheating temperature in this case has not to
be
too high, otherwise another mechanism of baryogenesis has to be
implemented in the model to explain the observed 
matter-over-antimatter asymmetry of the Universe.

Note in passing that in principle additional operators in quark sector
can also contribute in a similar way to the generation of the baryon
asymmetry, but these operators are proportional to higher powers of
$H^\dagger{}H/\Lambda^2$ and the generic analysis is rather involved.

We conclude this Section by the statement that with higher order operators
in Higgs-inflation there are possibilities to  explain both the
neutrino oscillations and baryon asymmetry of the Universe. 


\section{Higher dimensional operators in the $\nu$MSM and dark matter
  sterile neutrinos}
\label{Sec:nuMSM}

DM problem needs introduction of some new particles in the SM. A
minimal example of such physics is $\nu$MSM
\cite{Asaka:2005pn,Boyarsky:2009ix}, the Standard Model extension with
three sterile neutrinos $N_I$, $I=1,2,3$, lighter than electroweak
bosons.  To the SM Lagrangian the following terms are added
\begin{equation}
\label{nuMSM-lagrangian}
\delta L^{\nu\,{\mathrm{MSM}}}=i\bar N_I \partial_\mu \gamma^\mu N_I + \half \,
M_I \bar N_I^c N_I - f_{\alpha I}\bar L_\alpha N_I \tilde H + \mathrm{h.c.}
\end{equation}
The lightest sterile neutrino, say mostly $N_1$, comprises dark matter
while the two heavier sterile neutrinos due to nonzero Yukawa
couplings $f_{\alpha I}\ll 1$ provide active neutrino with masses.
Sterile neutrinos can contribute to the dark matter and lepton asymmetry
generation.  One of the important difficulties in the plain $\nu$MSM
setup is the generation of correct abundance of the DM sterile
neutrino (see \cite{Boyarsky:2009ix} for a review), which requires
very severe fine tuning in sterile neutrino sector \cite{Roy:2010xq} 
or a special source of dark matter neutrino in the early Universe,
e.g. decaying inflaton \cite{Shaposhnikov:2006xi,Bezrukov:2009yw}.  We
consider in this Section whether the higher order operators
(\ref{general-operators}) can change the situation.

The sterile neutrino has renormalizable Yukawa interactions
\eqref{nuMSM-lagrangian} with the fields of the SM.  These
interactions inevitably give rise to the decays of $N$ such as $N\to3\nu$
and $N\to\gamma\nu$, and to its production in the early Universe.  The
dark matter candidate in the form of sterile neutrino was introduced
in \cite{Dodelson:1993je} (for a review see \cite{Boyarsky:2009ix} and
references therein).  The analysis of the non-resonant production of
neutrinos and the X-ray constrains puts an upper limit on its mass,
$M<4$ keV, which happens to be in conflict with the lower limits on
from Ly-$\alpha$ analysis $M>8$ keV.\footnote{The resonant production
of DM sterile neutrinos \cite{Shi:1998km,Laine:2008pg}, which occurs
in the presence of lepton asymmetry, escapes these constraints
\cite{Boyarsky:2008mt}.}

We will see, that the higher-dimensional operators can change the
situation and lead to efficient production of this DM candidate,
consistent with other constraints.  Still, even in this case the mass
of sterile neutrino should be relatively small, as is argued below.


\subsection{Stability at cosmological time-scales}

The assumption that \emph{all} higher dimensional operators are
included is very powerful as it allows to conclude \emph{generally}
that any DM candidates have to be unstable (though very long-lived)
and relatively light, so that the best strategy for their search is
searches for emerging in dark matter decay photons of MeV or lower
energies. Considering $\nu$MSM as an example, the argument above goes
as follows.

Even if all couplings $f_{\alpha\,1}$ in \eqref{nuMSM-lagrangian} are
zero, they will be generated by five-dimensional operators,
proportional to $b_{L_\alpha}$ in (\ref{general-operators}).  The use
of equations of motion for $N$ leads to an estimate of
\emph{effective} Yukawa coupling
\be
  \label{est-Yuk}
  f_{\alpha\, 1} \sim b_{L_\alpha}\,\frac{M_1}{\Lambda}
  \;.
\ee
So, $N_1$ is unstable due to higher dimensional operators, with the
$\gamma\nu$ partial width of the order
\be
  \Gamma_{N_1\to \gamma\nu}
  \sim \frac{9\,b_{L_\alpha}^2 \alpha G_F^2}{512 \pi^4}\frac{v^2
    M_1^5}{\Lambda^2}
  \;.
\ee
Taking a constraint from EGRET data
$\tau_{\gamma\nu} \gsim \unit[10^{27}]{s}$\cite{Bertone:2007aw}, we obtain that $M_1 \lsim \unit[200]{MeV}$
for $b_{L_\alpha}\sim 1$ and $\Lambda=M_P$.  Similar bounds follow from
five-dimensional operators like the second term in 
\eqref{general-operators-sterile} 
where derivatives act on
lepton doublets.  With smaller scale $\Lambda=M_P/\xi$ instead of
$M_P$ in this term (or larger than \eqref{est-Yuk} Yukawa
couplings), the upper limit on sterile neutrino mass becomes stronger,
\[
  M_1\lesssim\unit[4]{MeV}
  \;.
\]
It is also worth mentioning that light dark matter sterile neutrino
\emph{can not help to explain the active neutrino oscillations,} as it
contributes only a tiny amount to sterile neutrino mass via seesaw
mechanism in \eqref{nuMSM-lagrangian}; to gain active neutrino masses,
heavier unstable neutrinos are required, which however can be searched
for in particle physics experiments \cite{Gorbunov:2007ak}.


\subsection{Dark matter production}

The sterile neutrino $N_1$ can be generated from the dimension 5
operator in (\ref{general-operators}) during the hot stage and at the
preheating stage.

\paragraph{Production at hot stage.} The coupling, proportional to
$\beta_N$ in (\ref{general-operators}), produces sterile neutrinos in
the primordial plasma at hot stage of the Universe evolution.  If
stable at cosmological time scales this sterile neutrino is a viable
candidate for dark matter.  If it never came to equilibrium, its mass
is limited from above at a given $\beta_N/\Lambda$ by imposing
$\Omega_N\lesssim \Omega_{DM}$.  This limit can be read off from
eq.~(5.20) of ref.~\cite{Bezrukov:2008ut} after multiplying the
l.h.s.\ of that equation by $4\Lambda^2/M_P^2$ and choosing
$T_r=\Lambda$ and $\beta=\beta_N$.  This gives
\[
  M_N \lesssim \unit[215]{keV}\, \frac{\Lambda}{\beta_N^2\,M_P}\;.
\]
Equality in the equation above saturates if sterile neutrinos comprise
right amount of the dark matter.

We observe that with $\Lambda\sim M_P$ and $\beta_N\sim 1$ the dark
matter neutrino has mass in sub-MeV region.  The lower the scale
$\Lambda$, the lighter the dark matter neutrino.  With low energy
cutoff $\Lambda=M_P/\xi$ one obtains
\[
M_N\lesssim 10\times \l \frac{0.03}{\beta_N}\r^2~\unit{keV}\;.
\]
Neutrino average momentum is close to the thermal average momentum in
the primordial plasma, so the constraints from Ly-$\alpha$ discussed
in \cite{Boyarsky:2009ix} are relevant in our case either.  Thus, for
the cutoff $\Lambda=M_P/\xi$ the viable dark matter implies
$\beta_N\lesssim 0.03$, sterile neutrino might form warm dark
matter.  If $\beta_N\sim 1$ and $\Lambda=M_P/\xi$, stable at
cosmological time scales sterile neutrinos contribute to the \emph{hot
  dark matter} component, and hence can not solve the dark matter
problem.   With $\beta_N\sim 1$, $\Lambda=M_P$ the sterile neutrinos
are cold and can solve the problem.  Note, that a particular value of
dark matter mass $M_N$ should be corrected if higher order operators
change somewhat the reheating in the model, as we discuss in Section
\ref{subsec:neutr-rehe-via}.

However, we will see now, that at the preheating stage the neutrino 
production is much more active.

\paragraph{Dark Matter generation during preheating with  dimension-5
   operator suppressed by {\bf$\Lambda(h)$.}}
\label{sec:DMlambdagen}
Let us consider the sterile neutrino production by the same dimension
5 operator as that in previous Section, suppressed by the
Higgs-dependent scale, corresponding to the gravity-scalar cutoff
\eqref{scalescalar},
\begin{equation}
  \label{eq:1}
  \cL_{\mathrm{int}} = \beta_N \frac{H^\dagger H}{2\Lambda} \bar{N^c}N
  = \frac{\beta_N}{4}\frac{h^2}{\Lambda(h)} \bar{N^c}N
  \;.
\end{equation}
Calculation proceeds along the same lines as in Section
\ref{subsec:neutr-rehe-via}.  The sterile neutrino density at late
times is given by eq.~\eqref{eq:11} with obvious replacement
$\beta_L\to \beta_N$.

Dividing this density 
by the entropy at reheating we get for sterile neutrino abundance
\begin{multline}
  \label{eq:12}
  \Delta_N(X) =
  \frac{90}{g_*4\pi^2}
  {\beta_N^2} \frac{\sqrt{\lambda}}{2\sqrt{3}\pi^2}
  \xi\frac{\Xcr^3}{T_r^3}
  \\
  =
  \frac{90}{g_*4\pi^2}
  {\beta_N^2} \frac{\sqrt{\lambda}}{\sqrt{3}\pi^2}
  \frac{2^{3/2}}{3^{3/2}\xi^2}\frac{M_P^3}{T_r^3}
  ={1.7\times10^{-4}}\beta_N^2\sqrt{\frac{\lambda}{0.25}}
  \frac{1}{\xi^2}\frac{M_P^3}{T_r^3}
\end{multline}
(seemingly large $\xi$ suppression is not there, as it is compensated
by two extra powers of $M_p/T_r$).
Thus, we get for $\Omega_{DM}=0.223$
\begin{equation}
  \label{result-non-thermal}
  \frac{\Omega_N}{\Omega_{DM}} =
  \frac{\beta_N^2\;M_N}{\unit[2.4]{keV}}
  \,\sqrt{\frac{\lambda}{0.25}}
  \frac{1}{\xi^2}\frac{M_P^3}{T_r^3}
  = \beta_N^2\frac{M_N}{\unit[1.3\times10^{-4}]{keV}}
  \,\sqrt{\frac{0.25}{\lambda}}
  \left(\frac{\unit[1.1\times10^{14}]{GeV}}{T_r}\right)^3
  \;.
\end{equation}
We can see, that if the constant $\beta_N>10^{-2}$, the sterile
neutrino DM should be lighter than keV, which violates the structure
formation bounds.  Thus, if $\beta_N\sim 1$  it is a problem to
generate DM sterile neutrino with the higher dimensional operators;
however with  hierarchy $\beta_N\lesssim 10^{-2}$ it is still
possible.  Obviously, the case of gauge cutoff leads to even larger
production.

The case of the Planck mass cutoff (\ref{scalePlanck}) coincides with
the analysis in \cite{Bezrukov:2008ut}, which also allows successful
generation of the DM density
\begin{equation}
  \label{result-non-thermal-Mpl}
  \frac{\Omega_N}{\Omega_{DM}} =
  \beta_N^2 \frac{M_N}{\unit[10^4]{keV}} \l
  \frac{0.25}{\lambda}\r
  \left(\frac{\unit[1.1\times10^{14}]{GeV}}{T_r}\right)
  \;.
\end{equation}
In this case the sterile neutrino should be rather heavy, and
additional care is needed to assure its stability (see previous
section).

\section{Summary and conclusions}

We have analyzed various effects that can appear from the higher order
operators in Higgs inflationary theory.  In the previous analysis
\cite{Bezrukov:2008ut} we used additional operators suppressed by a
constant (field independent) cutoff, which lead to the conclusion that
either the inflationary properties of the theory are spoiled, or no
significant generation of any chemical potentials is possible at
preheating due to these operators.

In this paper we analyzed a wider class of operators, suppressed by
the Higgs dependent cut-off, which is essential to sustain good
properties of the theory \cite{Bezrukov:2010jz}.  The key observation
here is that now the cut-off can be lowered significantly at present
time and at reheating without spoiling the inflationary properties.
As a result, with addition of properly chosen higher dimensional
operators one can explain many effects.  With proper tuning of the
coefficients in front of these operators, one can explain neutrino
masses, obtain proton decay, generate baryon asymmetry and sterile
neutrino DM (within $\nu$MSM framework).  However, all this is only
possible with rather specific choices of the hierarchies between the
higher dimensional operators, and a generic prediction seems to be
impossible.

The work is supported in part by the grant of the President of the
Russian Federation NS-5525.2010.2 (government contract
02.740.11.0244).  D.G.  thanks ITPP EPFL for hospitality. The work of
F.B. is partially supported by the Humboldt foundation.  The work of
D.G.  is supported in part by FAE program (government contract
$\Pi$520), RFBR grants \foreignlanguage{russian}{11-02-92108-ЯФ\_а,
  11-02-01528-а}, and by SCOPES program.
The work of M.S. is supported by the
Swiss National Science Foundation.

\label{summary}

\appendix

\section{Generation of fermions by varying mass term, scalar gravity
  sector cutoff case}
\label{app:nugen_scalargravity}

Let us analyze the generation of fermions by the operators
(\ref{general-operators-violating}) or
(\ref{general-operators-sterile}).  In the presence of the time
varying Higgs field background this leads to the Hamiltonian for the
fermions 
\[
  \hat H_{\mathrm{int}} \equiv \int d^3{\mathbf{x}}\, m\l t\r \bar\Psi \Psi
\]
with the time varying mass (\ref{eq:3}).  The analysis of the
generation of the fermions, while the amount of generated particles is
small (occupation numbers much smaller than unity) can be done
perturbatively and closely follows appendix F of
\cite{Bezrukov:2008ut}).
We treat the constant mass
$m_0={\beta_L}\sqrt{\frac{3}{\lambda}}\omega/4$ as a part of the free
Hamiltonian.
Then the number density at the moment t
\[
  n_k\l t \r = \int_0^t (m(t')-m_0) \e^{2ik_0t'} dt'
               \int _0^t (m(t'')-m_0) \e^{-2ik_0t''} d t''
  \;.
\]
and total particle number
\[
  n(t)=\int \frac{d^3\bk}{(2\pi)^3}n_k(t)
  \;.
\]
Integration is the same as in \cite{Bezrukov:2008ut}, leading to
\begin{equation}
  n_\bk( t_l) = \frac{1}{16}\frac{\sin^2\left( 2l\frac{\pi
      k_0}{\omega}\right)  }
  {\sin^2\left(\frac{\pi k_0}{\omega}\right)}
  L^2
  \;,
\end{equation}
where the constant is slightly changed:
\begin{equation}
  \label{eq:6}
  L={\beta_L}\sqrt{\frac{\lambda}{3}}\frac{1}{4\omega}X^2
  \cdot \frac{\omega}{k_0}
  \left\{
   \frac{1}{k_0+\omega} \sin\left[ 4\pi \left( \frac{k_0}{\omega}
      +1\right) \epsilon \right]
   - \frac{1}{k_0-\omega}
   \sin\left[4\pi \left( \frac{k_0}{\omega}-1\right) \epsilon
   \right]
  \right\}
  \;,
\end{equation}
where we removed $2k_0$ from the denominator and changed $3\lambda
X^2\to{\beta_L}\sqrt{\frac{\lambda}{3}}\frac{1}{4\omega}X^2$ (compare
(\ref{eq:3}) and
(C.1)
of \cite{Bezrukov:2008ut}); the
ratio $\omega/k_0$ should always stay, because it is the ratio of
the source frequency and the energy $k_0$ in the oscillating exponent.
The parameter $\epsilon$ is defined as
\begin{equation}
  \label{eq:7}
  \sin(2\pi\epsilon) \equiv
  \sqrt{\frac{3}{\lambda}}\frac{\omega}{X}
  \;.
\end{equation}
At large times the ratio leads to the sum of delta functions over
momenta, and integration over momenta gives
\begin{equation}
  \label{eq:8}
  \frac{n\left(t \right)}{ t} =
  {\beta_L^2} \frac{\lambda}{48} X^4 \frac{1}{(2\pi)^3}
  \sum_{l=2}^\infty
  \sqrt{1-\frac{1}{l^2}}
  \cdot
  \left(
   \frac{1}{l+1} \sin\left[4\pi \left(l+1\right)\epsilon
   \right]
   - \frac{1}{l-1}
   \sin\left[4\pi \left(l-1\right)\epsilon
   \right]
  \right)^2
  \;.
\end{equation}
Summation for small $\epsilon$ can be performed using the integral
(cf.\ appendix
E
of \cite{Bezrukov:2008ut})
\begin{equation}
  \label{eq:9}
  S_3 = \int_0^\infty (4\pi\epsilon)^3
  \frac{du}{u^4}
  \left( 2 u \cos u -2 \sin u \right)^2
  = \frac{2\pi}{3} (4\pi \epsilon)^3
  \;.
\end{equation}
This finally yields the generation rate (we reintroduced also the
expansion of the Universe here)
\begin{equation}
  \label{eq:10}
  \frac{d}{dt}(a^3 n) = a^3
  {\beta_L^2}\sqrt{\frac{3}{\lambda}}\frac{1}{24\pi^2}
  X\omega^3
  \;.
\end{equation}
Integrating this during the matter dominated expansion
(3.4)
from \cite{Bezrukov:2008ut} we obtain the neutrino density 
(saturated at late times)
\begin{equation}
  \label{eq:11}
  n = {\beta_L^2} \frac{\sqrt{\lambda}}{16\sqrt{3}\pi^2}
  \xi\Xcr^3
  \;.
\end{equation}
Note, that this does not depend on time, because rate (\ref{eq:7}) is
proportional to $X\propto t^{-1}$.  This formula should be compared
to
\(
  n_N(X) =
  \frac{\beta^2\sqrt{\lambda}}{48\sqrt{2}\pi\xi}\Xcr^2 X
\)
from Section 5.2.2 of \cite{Bezrukov:2008ut}.

Thus, the total active neutrino density is trice the amount of
(\ref{eq:11}).  To check, whether this leads to preheating, we should
compare the energy density in the neutrinos to those in the inflaton. 
To do this, we need energy density instead of the number density. 
Just doing the computation using occupation numbers (\ref{eq:5}) leads
to divergent result.  This is caused by  discontinuous approximation
to mass formula (\ref{eq:3}). However, for the estimate purposes we
can just consider, that the average energy of created neutrinos is
still given by the value in appendix C of \cite{Bezrukov:2008ut},
i.e.\footnote{One can avoid the divergence in the calculation of
energy by smoothing the effective mass \eqref{eq:3}, say, as follows
  \begin{equation*}
   m_0-m(t) =
    \frac{\frac{\beta_L}{2\sqrt{2}}\sqrt{\frac{3}{2\lambda}}}
           {\cosh\left(\sqrt{\nicefrac{2\lambda}{3}}X\cdot t\right)}
    \;.
  \end{equation*}
  Then, instead of (\ref{eq:5}) we get
  \begin{equation*}
    n_k(t) =
    \frac{9\pi\beta_L^2\omega^2}
         {32\lambda^2X^2\cosh^2\left(\sqrt{\frac{3}{2\lambda}}\frac{\pi k_0}{X}\right)}
    \frac{\sin^2\left( 2l\frac{\pi k_0}{\omega}\right)  }
    {\sin^2\left(\frac{\pi k_0}{\omega}\right)  }
    \;.
  \end{equation*}
  Here we obviously get an exponential energy cut-off with typical scale
  (\ref{eq:14}).}
\begin{equation}
  \label{eq:14}
  E\sim\frac\omega{4\pi\epsilon} =
  \frac{1}{2}\,\sqrt{\frac{\lambda}{3}} X
  \;.
\end{equation}
Then we have for the energy density
\begin{equation}
  \label{eq:15}
  \rho \sim \frac{\lambda \beta_L^2}{96\,\pi^2}\xi\Xcr^3 X
  \;.
\end{equation}
Equating this with the inflaton energy density
(3.7)
in \cite{Bezrukov:2008ut} we get
\begin{equation}
  \label{eq:16}
  \frac{\lambda \beta_L^2}{96\,\pi^2}\xi\Xcr^3 X = \frac{\lambda}{4}\Xcr^2 X^2
  \;.
\end{equation}
Thus, the reheating (half energy in the relativistic active neutrinos)
happens at
\begin{equation}
  \label{eq:17}
  X \sim \frac{\beta_L^2}{24\,\pi^2}\xi\Xcr
  \;.
\end{equation}
This is parametrically earlier, than (3.10)
in
\cite{Bezrukov:2008ut}.  Thus, reheating temperature is
\begin{equation}
  \label{eq:18}
  T_r \sim \left(\frac{10\,\lambda}{27\,g_*}\right)^{1/4}
  \frac{\beta_L}{2\pi\sqrt{2\pi}}\frac{M_P}{\sqrt{\xi}}
  \;.
\end{equation}
This is parametrically larger than the
temperature of reheating due to renormalizable operators
\cite{Bezrukov:2008ut},
\[
T_r^{\mathrm{renorm}}\propto \frac{M_P}{\xi}\;,
\]
but for $\beta_L\sim 1$ it gives the numerical estimate of the same order,
\[
T_r \simeq 1.7\times \beta_L \times
  10^{14} \,\unit{GeV} \sim T_r^{\mathrm{renorm}}\simeq 10^{14}\,\unit{GeV}\;.
\]

Above we neglected the Pauli blocking effect for the fermions.
Indeed, with estimates \eqref{eq:11} and \eqref{eq:14} we get for the
occupation numbers at reheating
\[
  n_\bk\frac{n}{E^3}
  \simeq 10^3\times \beta_L \left(\frac{\Xcr}{X_r}\right)^3
  \;.
\]
This reaches one (breakdown of perturbative approximation because of
Pauli blocking effects) for $\beta_L\sim1$ at the higher reheating
temperature in \cite{Bezrukov:2008ut} (see formula (3.21) there).
So, the Pauli blocking effects were not really relevant for this calculation.

\bibliographystyle{JCAP-hyper}
\bibliography{all,misc}

\end{document}